\documentclass{SciPost}

\hypersetup{
    colorlinks,
    linkcolor={red!50!black},
    citecolor={blue!50!black},
    urlcolor={blue!80!black}
}

\usepackage[bitstream-charter]{mathdesign}
\DeclareSymbolFont{usualmathcal}{OMS}{cmsy}{m}{n}
\DeclareSymbolFontAlphabet{\mathcal}{usualmathcal}

\fancypagestyle{SPstyle}{
\fancyhf{}
\lhead{\colorbox{scipostblue}{\bf \color{white} ~SciPost Physics }}
\rhead{{\bf \color{scipostdeepblue} ~Submission }}

\fancyfoot[C]{\textbf{\thepage}}
}

\begin{document}

\pagestyle{SPstyle}

\begin{center}{\Large \textbf{\color{scipostdeepblue}{
Quantized resonant tunneling effect in Josephson junctions with ferromagnetic bilayers\\
}}}\end{center}

\begin{center}\textbf{
Hao Meng\textsuperscript{1$\star$},
Lei Cai\textsuperscript{1},
Xiuqiang Wu\textsuperscript{2,3},
Xianghe Zhao\textsuperscript{1},
Jia Xu\textsuperscript{1} and
Guanqi Wang\textsuperscript{1}
}\end{center}

\begin{center}
{\bf 1} School of Physics and Telecommunication Engineering, Shaanxi University of Technology, Hanzhong 723001, China
\\
{\bf 2} School of Optical and Electronic Information, Suzhou City University, Suzhou 215104, China
\\
{\bf 3} Suzhou Key Laboratory of Biophotonics, Suzhou 215104, China
\\[\baselineskip]
$\star$ \href{mailto:email1}{\small menghao2021@163.com}
\end{center}

  \section*{\color{scipostdeepblue}{Abstract}}
  \textbf{\boldmath{%
   We study the Josephson effect in one-dimensional SF$_1$F$_2$S junctions, which consist of conventional s-wave superconductors (S) connected by two ferromagnetic layers (F$_1$ and F$_2$). At low temperatures, the potential barrier at the F$_1$/F$_2$ interface can induce a quantized resonant tunneling effect. This effect not only modifies the amplitude of the critical current but also affects the phase of the Josephson current. As the exchange fields ($h_1$, $h_2$) and thicknesses ($d_1$, $d_2$) of the F$_1$ and F$_2$ layers vary, the critical current displays periodic resonance peaks. These peaks occur under the quantization conditions $Q_{1(2)} d_{1(2)} = \left(n_{1(2)} + 1/2\right) \pi$, where $Q_{1(2)} = 2h_{1(2)}/(\hbar v_F)$ is the center-of-mass momentum carried by Cooper pairs, with $v_F$ being the Fermi velocity, and $n_{1(2)} = 0, 1, 2, \cdots$. It can be inferred that the potential barrier suppresses the transport of spin-singlet pairs while allowing spin-triplet pairs with zero spin projection along the magnetization axis to pass through. As these spin-triplet pairs traverse the F$_1$ and F$_2$ layers, the total phase they acquire determines the ground state of the Josephson junction. At the resonance peaks, the Josephson current primarily arises from the first harmonic in both the parallel and antiparallel magnetization configurations. However, in perpendicular configurations, the second harmonic becomes more significant. In scenarios where both ferromagnetic layers have identical exchange fields and thicknesses, the potential barrier selectively suppresses the current in the 0-state while allowing it to persist in the $\pi$-state for parallel configurations. Conversely, in antiparallel configurations, the current in the 0-state is consistently preserved.
}}

\vspace{\baselineskip}

\noindent\textcolor{white!90!black}{%
\fbox{\parbox{0.975\linewidth}{%
\textcolor{white!40!black}{\begin{tabular}{lr}%
  \begin{minipage}{0.6\textwidth}%
    {\small Copyright attribution to authors. \newline
    This work is a submission to SciPost Physics. \newline
    License information to appear upon publication. \newline
    Publication information to appear upon publication.}
  \end{minipage} & \begin{minipage}{0.4\textwidth}
    {\small Received Date \newline Accepted Date \newline Published Date}%
  \end{minipage}
\end{tabular}}
}}
}


\vspace{10pt}
\noindent\rule{\textwidth}{1pt}
\tableofcontents
\noindent\rule{\textwidth}{1pt}
\vspace{10pt}


     \section{Introduction}
     \label{Sec1}
     The interplay between superconductivity and ferromagnetism has attracted significant interest because of its potential to produce fascinating physical phenomena and enhance advanced electronic devices~\cite{Golubov,Buzdin,Bergeret,Linder,Eschg,Eschrig,JLAVBa,YMShuk2022, IVBAMB2022}. The distinctive physical properties of heterostructures composed of superconductors (S) and ferromagnets (F) present a promising research area for future superconducting spintronics and quantum computing~\cite{ASMSVM2022,RCai,NOBirge2024}. It is well established that ferromagnetism and spin-singlet superconductivity represent two antagonistic orders: ferromagnetism aligns electron spins parallel to one another. In contrast, spin-singlet Cooper pairs consist of electrons with opposite spins. The two competing orders not only give rise to unconventional types of Cooper pairs but also realize tunable Josephson junctions.

     In hybrid S/F structures with homogeneous magnetization, the Cooper pairs \((\uparrow\downarrow - \downarrow\uparrow)\) can penetrate into the ferromagnetic region, acquiring a finite center-of-mass momentum \(Q=2h/({\hbar}v_F)\) due to the exchange splitting in the F~\cite{Eschrig}. Here, \(h\) represents the exchange field, and \(v_F\) denotes the Fermi velocity. This momentum shift drives the formation of spin-singlet pairs and spin-triplet pairs with zero spin projection along the magnetization axis: \((\uparrow\downarrow)e^{iQR}-(\downarrow\uparrow)e^{-iQR}=(\uparrow\downarrow-\downarrow\uparrow)\cos(QR)+i(\uparrow\downarrow+\downarrow\uparrow)\sin(QR)\), where \(R\) indicates the position relative to the S/F interface~\cite{Eschg,Eschrig}. Both the spin-singlet pairs \((\uparrow\downarrow - \downarrow\uparrow)\) and the spin-triplet pairs \((\uparrow\downarrow + \downarrow\uparrow)\) oscillate with the variable \(QR\) and have a short penetration depth into the F region. In a uniform ferromagnetic Josephson junction (SFS), the oscillation of the spin-singlet pairs within the F region gives rise to a phenomenon known as the 0-\(\pi\) transition. In this context, the ground state can either be a 0-junction with equal superconducting phases or a \(\pi\)-junction in which the phases differ by \(\pi\)~\cite{Buzdin,Eschg}. This transition manifests as a sign reversal in the critical current with variations in temperature or ferromagnetic thickness (see~\cite{Buzdin} and the references cited therein).

     On the other hand, a nonuniform ferromagnet can generate equal-spin triplet pairs, which consist of either spin-up (\(\uparrow\uparrow\)) or spin-down (\(\downarrow\downarrow\)) electrons that belong to the same spin band. These triplet pairs are immune to the exchange field and can travel long distances within the F region~\cite{Buzdin,Bergeret,Linder,Eschg,Eschrig,JLAVBa,FSBAF2001,AKRIS2001, KHPHB2007,KHOTVPHB2008,KHOTV2009,KHaMAl2016,CTWKH2018}. In Josephson junctions with nonuniform magnetization, the presence of equal-spin triplet pairs in the F region results in a long-range supercurrent~\cite{Bergeret,Linder,Eschg,Eschrig,MEsJKo2003,MHoAIB2007, YAsYTa2007, MEsTLof2008,HMeXWu2013,RSKSTB2006,JWAJDS2010,TSKMAK2010,CKlTSK2012,WMMWPP2016,NBaJWA2014,KHMARS2022,HMXWYR2022,DSMaSM2022}. Additionally, the Josephson junctions containing a ferromagnetic spin valve (SF\(_1\)F\(_2\)S) display a variety of intriguing phenomena~\cite{YMBlanter,Bell,Crouzy,Robinson,Trifunovic,LTrifunov,Melniko,Knezevic,Richard,SHikin,Baek1,Qader,Baek2,Gingrich, HMeng,Meng,BMNiedzielski}. The anharmonic current-phase relation can be expressed as \(I(\phi) = I_{1}\sin(\phi) + I_{2}\sin(2\phi) + \cdots\)\cite{Golubov}, where \(I_n\) denotes the supercurrent amplitude corresponding to the \(n\)th Josephson harmonic, illustrating the coherent transfer of \(n\) Cooper pairs~\cite{LTrifunov}, and \(\phi\) represents the superconducting phase difference. In highly asymmetric SF\(_1\)F\(_2\)S junctions, a significant second harmonic current (\(I_2 \gg I_1\)) arises at low temperatures~\cite{Trifunovic, LTrifunov, Richard, HMeng, Meng}.

     When an insulating potential barrier exists at the interface between the F\(_1\) and F\(_2\) layers, it results in several interesting characteristics in the current. Previously, Bergeret \emph{et al.} predicted that at low temperatures, the exchange field could enhance the Josephson critical current in the SF\(_1\)F\(_2\)S junction when the magnetizations of the F\(_1\) and F\(_2\) layers are antiparallel. They showed that the critical current could exceed that observed in the similar Josephson junction without the exchange field~\cite{FSBerg2}. Following this, multiple research groups conducted detailed theoretical studies on this junction~\cite{Koshina,Krivoruchko,Chtchelkatchev,AAGolu,XiaoweiLi,Pajovic}. Subsequently, Bergeret \emph{et al.}'s prediction was confirmed experimentally~\cite{Robinson}. Recently, we demonstrated that in antiparallel magnetization configurations, the potential barrier at the F\(_1\)/F\(_2\) interface leads to significant oscillations in the critical current as a function of the exchange field and the thickness of the ferromagnetic layers. Notably, within a range of small exchange fields and thin ferromagnetic layers, the magnetism causes the critical current to increase~\cite{HMengYJE}. Meanwhile, Nikoli\'{c} \emph{et al.} discovered new interference phenomena occurring at specific thicknesses of ferromagnetic layers in the SF\(_1\)F\(_2\)S junctions with perpendicular magnetizations~\cite{DNikoli}.

     However, three urgent questions remain to be addressed: (i) In the SF\(_1\)F\(_2\)S junctions with antiparallel magnetizations, what conditions must the exchange field and ferromagnetic thickness satisfy to maximize the critical current? (ii) When the magnetizations in the bilayers are parallel and perpendicular, do the critical currents exhibit the same behaviors as those observed in the antiparallel configuration? (iii) When the magnetizations of the bilayer layers are perpendicular to each other, does the second harmonic current appear at the positions where the critical current reaches its maximum?

     The purpose of this paper is to address the three questions proposed above. We numerically solve the Bogoliubov--de Gennes (BdG) equations to calculate the Josephson current in a one-dimensional SF\(_1\)F\(_2\)S junction. For fully transparent F\(_1\)/F\(_2\) interfaces, the Josephson critical current oscillates continuously with increases in the exchange fields and thicknesses of the F\(_1\) and F\(_2\) layers, assuming both ferromagnets are in the same direction. This behavior results from the transport of spin-singlet pairs (\(\uparrow\downarrow - \downarrow\uparrow\)). Moreover, the potential barrier at the F\(_1\)/F\(_2\) interface induces a resonant tunneling effect at low temperatures. When the exchange fields and the ferromagnetic thicknesses meet the quantized tunneling conditions, the critical current displays periodic resonance peaks. We deduce that this resonant tunneling behavior arises from the coherent transmission of the spin-triplet pairs (\(\uparrow\downarrow + \downarrow\uparrow\)). Furthermore, the conditions for the occurrence of current resonance peaks remain unchanged in both perpendicular and antiparallel magnetization configurations. In contrast, in parallel and antiparallel configurations, the resonance current originates from the first harmonic, while in the perpendicular configurations, it arises from the second harmonic. Interestingly, a phase difference of \(\pi\) exists at the same current resonance peaks for the parallel and antiparallel configurations.

     This paper introduces three key innovations that distinguish it from previous works:

     (i) Our previous paper \cite{HMengYJE} focuses on cases where the F$_1$ and F$_2$ layers have identical exchange fields and thicknesses. It discussed how the critical current oscillates with variations in the exchange field and thickness for antiparallel magnetization configurations. However, it did not establish the necessary conditions for resonance peaks in the critical current nor clarify the underlying physical mechanisms driving the current oscillation. In contrast, this paper expands the criteria for current variation to include any exchange fields and thicknesses. The conditions for resonance peaks are identified as $Q_1 d_1 = (n_1 + 1/2)\pi$ and $Q_2 d_2 = (n_2 + 1/2)\pi$, where $Q_{1(2)}=2h_{1(2)}/(\hbar v_F)$ represents the center-of-mass momentum carried by Cooper pairs in the F$_{1(2)}$ layer, and $n_1$ and $n_2 = 0, 1, 2, \cdots$. Here, $v_F$ is the Fermi velocity, while $h_{1(2)}$ and $d_{1(2)}$ denote the exchange fields and thicknesses of the F$_{1(2)}$ layers, respectively. These behaviors arise from the resonant tunneling of spin-triplet pairs (\(\uparrow\downarrow + \downarrow\uparrow\)).

     (ii) The paper \cite{HMengYJE} calculates the critical current in parallel magnetization configurations and notes that the critical current amplitude decreases rapidly with increasing exchange field and thickness in three-dimensional structures. In this case, the current oscillation resulting from the 0-$\pi$ transition overlaps with the oscillation induced by resonant tunneling. Consequently, the resonance peaks in the current are obscured, making it challenging to identify the conditions necessary for resonant tunneling. In our current paper, we observe that the resonant tunneling effect becomes more pronounced in one-dimensional structures. When faced with sufficiently strong barriers, the current oscillation associated with the 0-$\pi$ transition is significantly suppressed, allowing the oscillation peaks related to resonant tunneling to stand out more clearly. This enhancement facilitates the identification of the conditions required for the occurrence of the resonant tunneling effect.

     (iii) Nikoli\'{c} \emph{et al.} indicate that in the three-dimensional SF$_1$F$_2$S junctions with perpendicular magnetizations, geometric interference occurs at specific ferromagnetic thicknesses, such as $d_1=d_2$, $d_2/3$, $d_2/5$, and so on, when the transparency of the F$_1$/F$_2$ interface is low~\cite{DNikoli}. They suggest that this interference arises from the first harmonic and is related to the multiple reflections that lead to the emergence of electron and hole quasiclassical trajectories with a canceled phase accumulation~\cite{DNikoli}. In contrast, the resonant tunneling effect discussed in this paper occurs within the one-dimensional junctions. This effect depends on both the ferromagnetic thicknesses and the exchange fields. Moreover, the potential barrier selectively filters out the first harmonic, allowing only the second harmonic to play a crucial role in the resonant tunneling process. Therefore, the resonant tunneling effect explored in this paper is different from the geometric resonances addressed in Ref.~\cite{DNikoli}.

     The paper is organized as follows: In Sec.~\ref{Sec2}, we present the model and the solution. In Sec.~\ref{Sec3}, we discuss the numerical results concerning the Josephson currents and their resonant tunneling effects. Finally, concluding remarks are given in Sec.~\ref{Sec4}.

     \section{Model and formula}
     \label{Sec2}

\begin{figure}[ptb]
        \centering
        \includegraphics[width=4.0in]{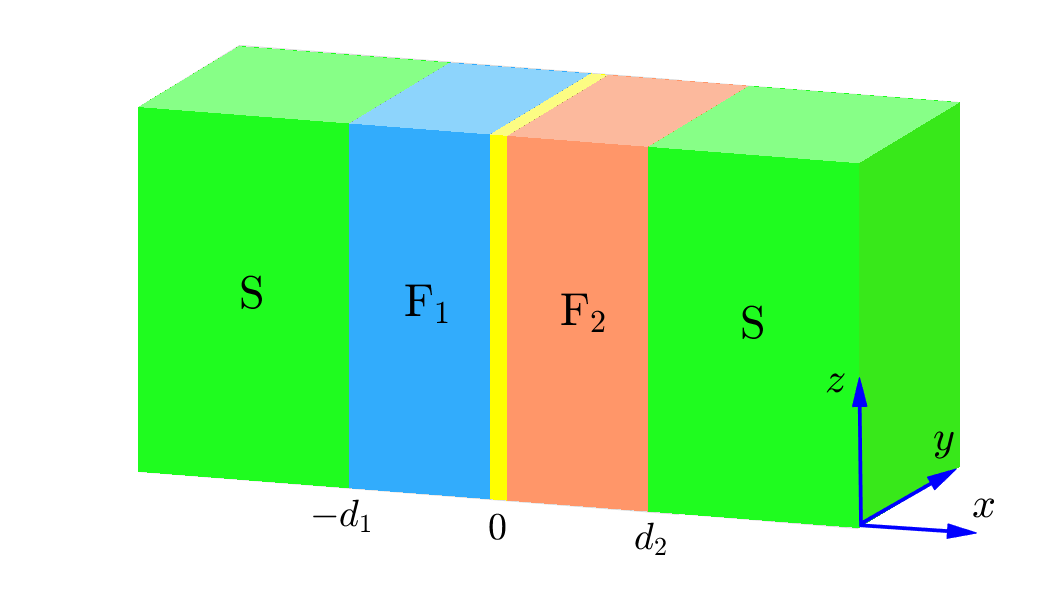}
        \caption{Schematic representation of the SF\(_{1}\)F\(_{2}\)S Josephson junction with a potential barrier at F\(_1\)/F\(_2\) interface. The thicknesses of F\(_1\) and F\(_2\) are denoted by \(d_1\) and \(d_2\), respectively.}
        \label{Fig1}
     \end{figure}

     As shown schematically in Fig.~\ref{Fig1}, we consider a one-dimensional SF\(_1\)F\(_2\)S Josephson junction, where a potential barrier resides at the interface between the F\(_1\) and F\(_2\) layers. The transport direction is along the \(x\)-axis. The effective Hamiltonian in the BCS mean-field framework can be expressed as follows~\cite{Buzdin, PGdeGennes}:
      \begin{align}
         H_{\rm {eff}}=&{\displaystyle\sum\limits_{\alpha,\beta}}\int{d}^3\mathbf{r}\left\{
         \hat{\psi}_{\alpha}^{\dagger}(\mathbf{\mathbf{r}})\left[H_{e}+U(\mathbf{r})\right]\hat{\psi}_{\alpha}(\mathbf{r})+\frac{1}{2}\left[(i\hat{\sigma}_{y})_{\alpha\beta}\Delta(\mathbf{r}
         )\hat{\psi}_{\alpha}^{\dagger}(\mathbf{r})\hat{\psi}_{\beta}^{\dagger}(\mathbf{r})+{\rm H.c.}\right]\right. \nonumber \\
         &-\left.\hat{\psi}_{\alpha}^{\dagger}(\mathbf{r})\left(\vec{h}\cdot\vec{\sigma}\right)_{\alpha\beta}\hat{\psi}_{\beta}(\mathbf{r})\right\} ,\label{HEFF}
      \end{align}
     where \( H_{e} = -\frac{\hbar^{2} \nabla^{2}}{2m} - E_{F} \) represents the quasiparticle kinetic energy relative to the Fermi energy \(E_F\). Here, \( \hat{\psi}^{\dagger}_{\alpha}(\mathbf{r}) \) and \( \hat{\psi}_{\alpha}(\mathbf{r}) \) denote the creation and annihilation operators with spin \(\alpha\). \( \vec{\sigma} = (\hat{\sigma}_{x}, \hat{\sigma}_{y}, \hat{\sigma}_{z}) \) is the vector of Pauli matrices. In this context, \( m \) is the effective mass of the quasiparticles in both the superconductors and the ferromagnets. We assume uniform Fermi energy \( E_F \) across all regions of the Josephson junction. The superconducting pair potential is spatially defined as \( \Delta(\mathbf{r}) = \Delta [ e^{i\phi/2} \Theta(-x-d_{1}) + e^{-i\phi/2} \Theta(x-d_{2}) ] \), where \(\Delta\) represents the bulk superconducting gap, \(\phi\) is the macroscopic phase difference between the two superconducting electrodes, and \(\Theta(x)\) denotes the Heaviside step function. This approximation is justified when the thickness of the superconducting layers far exceeds that of the ferromagnetic layers. The exchange field \(\vec{h}_{1}\) in the F\(_1\) layer aligns with the \(z\)-axis, while the exchange field \(\vec{h}_{2}\) in the F$_2$ layer is oriented at a polar angle \(\theta\) with the \(z\)-axis and an azimuthal angle \(\chi\) with the \(x\)-axis in the \(x\)-\(y\) plane, represented as \( \vec{h}_{2} = h_{2} (\sin\theta \cos\chi, \sin\theta \sin\chi, \cos\theta) \). This parametrization allows systematic control over the relative magnetization alignment (parallel, antiparallel, or canted) between the F\(_1\) and F\(_2\) layers. The F\(_1\)/F\(_2\) interface can be modeled by a spin-independent \( \delta \)-function potential barrier, described as \( U(\mathbf{r}) = V \delta(x)\). This idealized barrier simulates interfacial disorder or oxide layers.

     In this paper, we utilize the theoretical framework established in Ref.~\cite{HMengYJE}. To derive the BdG equations, we apply the Bogoliubov transformation \( \hat{\psi}_{\alpha}(\mathbf{r}) = \sum_{n} [ u_{n\alpha}(\mathbf{r}) \hat{\gamma}_{n} + v_{n\alpha}^{\ast}(\mathbf{r}) \hat{\gamma}_{n}^{\dag} ] \), where \( u_{n\alpha}(\mathbf{r}) \) and \( v_{n\alpha}(\mathbf{r}) \) are the electron and hole components of the quasiparticle wave function, and \(\hat{\gamma}_{n}^{\dag}\), \(\hat{\gamma}_{n}\) are creation, annihilation operators for quasiparticles. Using the expansion \( u_{n\alpha}(\mathbf{r}) = u_{k}^{\alpha} e^{ikx} \) and \( v_{n\alpha}(\mathbf{r}) = v_{k}^{\alpha} e^{ikx} \), we obtain the following BdG equations~\cite{PGdeGennes}:
     \begin{equation}
        \begin{pmatrix}
           \hat{H}_{0}+\hat{h} & i\hat{\sigma}_{y}\Delta(x)\\
           -i\hat{\sigma}_{y}\Delta^{\ast}(x) & -\hat{H}_{0}-\hat{h}^{\ast}
        \end{pmatrix}
        \begin{pmatrix}
           \hat{u}_{k}\\
           \hat{v}_{k}
        \end{pmatrix}
        =\epsilon
        \begin{pmatrix}
           \hat{u}_{k}\\
           \hat{v}_{k}
        \end{pmatrix},\label{BdG}
     \end{equation}
     where
      \[
      \hat{H}_{0}=
      \begin{pmatrix}
          \xi_{k}+V\delta(x) & 0\\
          0 & \xi_{k}+V\delta(x)
      \end{pmatrix},
      \text{and} \
      \hat{h}=
      \begin{pmatrix}
          -h_{z} & -h_{x}+ih_{y}\\
          -h_{x}-ih_{y} & h_{z}
      \end{pmatrix}.
      \]
      Here \( \xi_{k} = \frac{\hbar^{2} k^{2}}{2m} - E_{F} \), and the quasiparticle and quasihole wave functions are denoted by \( \hat{u}_{k} = (u_{k}^{\uparrow}\; u_{k}^{\downarrow})^{T} \) and \( \hat{v}_{k} = (v_{k}^{\uparrow}\; v_{k}^{\downarrow})^{T} \), respectively.

      The BdG equations (\ref{BdG}) can be solved for each superconducting electrode and each ferromagnetic layer. For a given energy \(\epsilon\) within the superconducting gap, we find the following plane-wave function in the left superconducting electrode:
      \begin{equation}
          \psi_{L}^{S}(x)= C_{1} \hat{\zeta}_{1} e^{-ik_{S}^{+}x} + C_{2} \hat{\zeta}_{2} e^{ik_{S}^{-}x}
            + C_{3} \hat{\zeta}_{3} e^{-ik_{S}^{+}x} + C_{4} \hat{\zeta}_{4} e^{ik_{S}^{-}x}, \label{functionSL}
      \end{equation}
      where \( k_{S}^{\pm} = k_{F} \sqrt{1 \pm i \sqrt{\Delta^{2} - \epsilon^{2}}/E_{F}} \) are the wave vectors for quasiparticles in the superconducting regions. $\hat{\zeta}_{1}=[1\;0\;0\;R_{1}e^{-i\phi/2}]^{T}$, $\hat{\zeta}_{2}=[1\;0\;0\;R_{2}e^{-i\phi/2}]^{T}$, $\hat{\zeta}_{3}=[0\;1\;-R_{1}e^{-i\phi/2}\;0]^{T}$, and $\hat{\zeta}_{4}=[0\;1\;-R_{2}e^{-i\phi/2}\;0]^{T}$ are the four basis wave functions of the left superconductor, where \( R_{1(2)}=(\epsilon\mp{i}\sqrt{\Delta^{2}-\epsilon^{2}})/\Delta \). Similarly, the wave function in the right superconducting electrode is
      \begin{equation}
          \psi_{R}^{S}(x) = D_{1} \hat{\eta}_{1} e^{ik_{S}^{+}x} + D_{2} \hat{\eta}_{2} e^{-ik_{S}^{-}x}
              + D_{3} \hat{\eta}_{3} e^{ik_{S}^{+}x} + D_{4} \hat{\eta}_{4} e^{-ik_{S}^{-}x}, \label{functionSR}
      \end{equation}
      where \( \hat{\eta}_{1}=[1\;0\;0\;R_{1}e^{i\phi/2}]^{T} \), \( \hat{\eta}_{2}=[1\;0\;0\;R_{2}e^{i\phi/2}]^{T} \), \( \hat{\eta}_{3}=[0\;1\;-R_{1}e^{i\phi/2}\;0]^{T} \), and \( \hat{\eta}_{4}=[0\;1\;-R_{2}e^{i\phi/2}\;0]^{T} \).

      The wave function in the F\(_{2}\) layer is given by
      \begin{align}
          \psi_{2}(x)&=(M_{1}e^{ik_{1}x}+M_{1}^{\prime}e^{-ik_{1}x})\hat{e}_{1}
             +(M_{2}e^{ik_{2}x}+M_{2}^{\prime}e^{-ik_{2}x})\hat{e}_{2}  \nonumber\\
            &+(M_{3}e^{ik_{3}x}+M_{3}^{\prime}e^{-ik_{3}x})\hat{e}_{3}
             +(M_{4}e^{ik_{4}x}+M_{4}^{\prime}e^{-ik_{4}x})\hat{e}_{4},\label{HM_wave}
          \end{align}
      where $\hat{e}_{1}=(\cos\frac{\theta}{2}\;\sin\frac{\theta}{2}e^{i\chi}\;0\;0)^{T}$, $\hat{e}_{2}=(-\sin\frac{\theta}{2}e^{-i\chi}\;\cos\frac{\theta}{2}\;0\;0)^{T}$, $\hat{e}_{3}=(0\;0\;\cos\frac{\theta}{2}\;\sin\frac{\theta}{2}e^{-i\chi})^{T}$, and $\hat{e}_{4}=(0\;0\;-\sin\frac{\theta}{2}e^{i\chi}\;\cos\frac{\theta}{2})^{T}$ are the basis wave functions in the F$_{2}$ layer. Moreover, $k_{1(2)}=k_{F}\sqrt{1+(\epsilon \pm{h_{2}})/E_{F}}$ and $k_{3(4)}=k_{F}\sqrt{1-(\epsilon\mp{h_{2}})/{E_{F}}}$ are the wave vectors of the quasiparticles in the F$_{2}$ layer. The corresponding wave function \(\psi_{1}(x)\) in the F\(_{1}\) layer can be derived from Eq.~(\ref{HM_wave}) by substituting \(h_{2} \rightarrow h_{1}\), \(\theta \rightarrow 0\), and \(\chi \rightarrow 0\).

      The wave functions and their first derivatives must satisfy continuity conditions at the interfaces of the SF\(_1\)F\(_2\)S junction:
        \begin{align}
          & \psi_{L}^{S}(-d_1) = \psi_{1}(-d_1), \frac{d\psi_{L}^{S}}{dx}\Bigr|_{x=-d_1} = \frac{d\psi_{1}}{dx}\Bigr|_{x=-d_1}, \label{boundary1}\\
          & \psi_{1}(0) = \psi_{2}(0), \frac{d\psi_{2}}{dx}\Bigr|_{x=0} - \frac{d\psi_{1}}{dx}\Bigr|_{x=0} = Zk_F \psi(0),  \label{boundary2}\\
          & \psi_{2}(d_2) = \psi_{R}^{S}(d_2), \frac{d\psi_{2}}{dx}\Bigr|_{x=d_2} = \frac{d\psi_{R}^{S}}{dx}\Bigr|_{x=d_2}. \label{condition3}
        \end{align}
      Here, the dimensionless parameter \(Z=2mV/(\hbar^{2}k_{F})\) characterizes the strength of the insulating potential barrier at the F\(_1\)/F\(_2\) interface.

      From these boundary conditions, we can establish 24 linear equations in the following form:
         \begin{equation}
            \Lambda\mathbf{X}=\mathbf{0},\label{linearEq}
         \end{equation}
      where \(\mathbf{X}\) contains 24 scattering coefficients, and \(\Lambda\) is a \(24 \times 24\) matrix. The solution of the characteristic equation
         \begin{equation}
            \det\left(\Lambda\right)=0   \label{characteristicEq}
         \end{equation}
      allows us to identify two Andreev bound-state solutions with energies \(E_{A\omega}\) (\(\omega = 1, 2\)). These discrete subgap states play a crucial role in Josephson transport within the short-junction limit, where the ferromagnetic thickness is much smaller than the superconducting coherence length (\(d_{1,2} \ll \xi_{S}\)). In this regime, the contributions from continuous quasiparticle states above the gap are negligible~\cite{PFBagwell, CWJBeenakker}. Therefore, the Josephson current in the SF\(_1\)F\(_2\)S junction is derived from the general formula
        \begin{equation}
           I(\phi)=\frac{2e}{\hbar}\frac{\partial\Omega}{\partial\phi},\label{current}
        \end{equation}
      where \(\Omega\) represents the phase-dependent thermodynamic potential. This potential is computed from the excitation spectrum using the formula~\cite{JBardeen,JCayssol}
         \begin{equation}
            \Omega=-2T\sum_{\omega}\ln\left[ 2\cosh\frac{E_{A\omega}(\phi)}{2T}\right], \label{potential}
         \end{equation}
      where the summation includes all positive Andreev energies (\(0 < E_{A\omega}(\phi) < \Delta\)). The parameters \(\Delta\), \(h_1\), \(h_2\), \(Z\), \(\theta\), and \(\chi\) are assumed to be equilibrium values that minimize the free energy of the SF\(_1\)F\(_2\)S junction~\cite{Buzdin-AdvPhys85}. For each phase difference \(\phi\), we numerically solve Eq.~(\ref{characteristicEq}) to determine two Andreev energies. From these Andreev energies, we can calculate the Josephson current \(I(\phi)\) using Eqs.~(\ref{current}) and (\ref{potential}). The critical current is then defined as \(I_{c} = \max_{\phi}|I(\phi)|\).

      \section{Results and discussions}
      \label{Sec3}

        \begin{figure}
           \centering
           \includegraphics[width=5.8in]{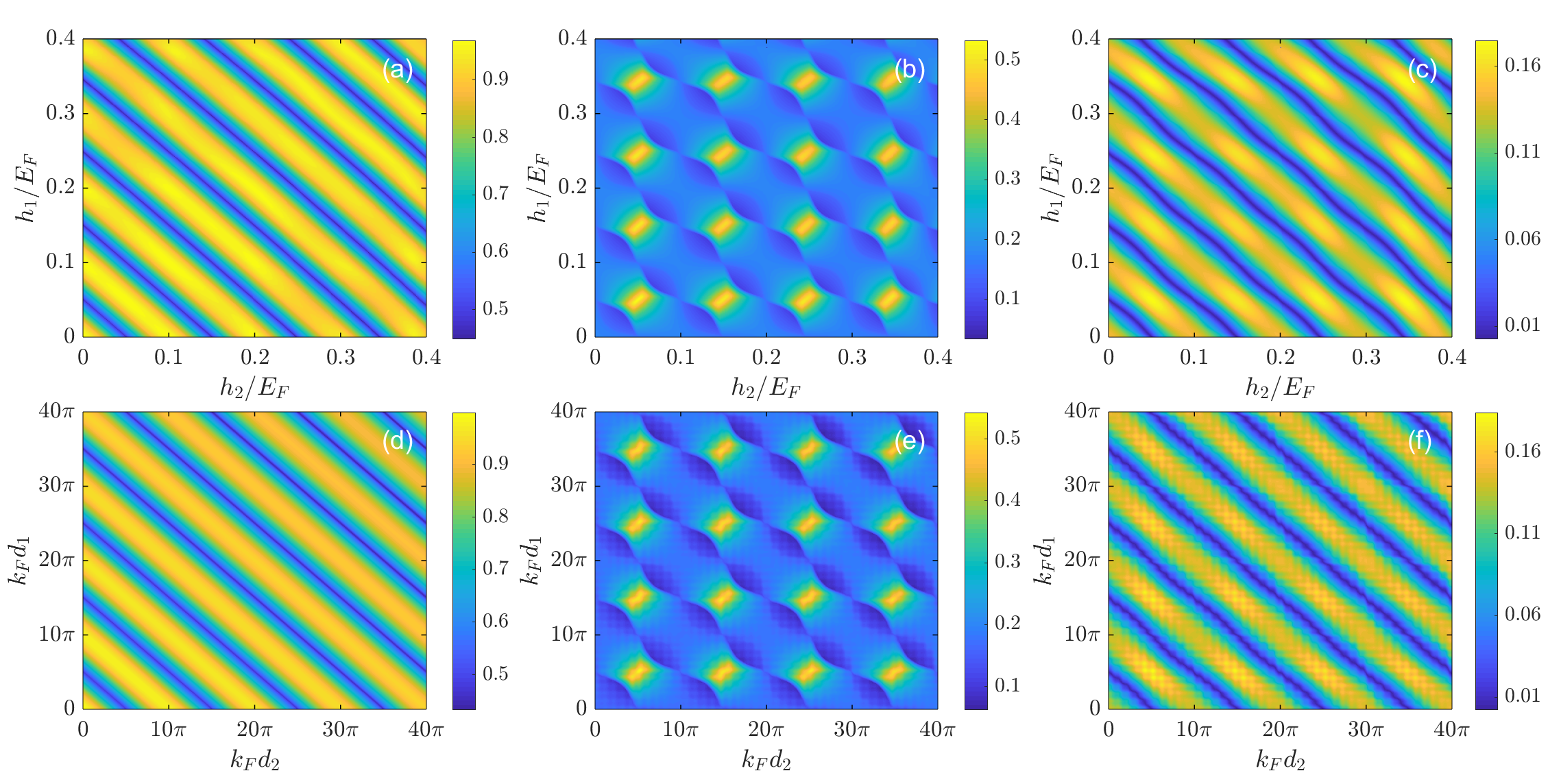} 
            \caption{The critical current \(I_c\) versus the exchange fields (\(h_{1}\), \(h_2\)) for the ferromagnetic thicknesses \(k_Fd_1=k_Fd_2=10\pi\) [(a), (b), and (c)], and \(I_c\) versus (\(d_1\), \(d_2\)) for \(h_1/E_F=h_2/E_F=0.1\) [(d), (e), and (f)]. The left column of graphs [(a) and (d)] corresponds to the barrier strength \(Z=0\), and the middle [(b) and (e)] and right [(c) and (f)] columns correspond to \(Z=3\). Additionally, the temperature is taken as \(T/\Delta=0\) for the left [(a) and (d)] and middle [(b) and (e)] columns and \(T/\Delta=0.4\) for the right [(c) and (f)] column. All panels are for the parallel magnetization configurations (\(\theta=0\) and \(\chi=0\)).}
            \label{Fig2}
        \end{figure}

       In our calculations, we use the superconducting gap \(\Delta\) as the unit of energy and set the Fermi energy to \(E_{F} = 1000 \Delta\), which is consistent with conventional metallic superconductors. Length scales are normalized using the inverse Fermi wave vector \(k_{F}^{-1}\), and exchange fields are scaled by the Fermi energy \(E_{F}\). Additionally, the normalized unit of the Josephson current is \( I_{0}=2e\Delta/\hbar \). Note that the approximation of the short Josephson junction (\(k_{F} d_1, k_{F} d_2 \ll 1000\)) is strictly satisfied in the presented calculations.

      \subsection{The Josephson current in configurations with parallel magnetization}

      We first investigate the Josephson current in the SF\(_1\)F\(_2\)S junction with parallel magnetization configurations. As illustrated in Figs.~\ref{Fig2}(a) and \ref{Fig2}(d), the critical current oscillates with the exchange fields (\(h_{1}\), \(h_{2}\)) and the ferromagnetic thicknesses (\(d_{1}\), \(d_{2}\)). This oscillation behavior appears as a continuous stripe pattern. The phase difference between two adjacent yellow stripes is \( \pi \). The blue stripe represents the 0-\(\pi\) transition point. At this point, the first harmonic current (\(I_{1}\)) vanishes, while the second harmonic current (\(I_{2}\)) appears completely. The underlying physical mechanisms for this current behavior can be explained as follows: When the Cooper pairs (\(\uparrow\downarrow-\downarrow\uparrow\)) enter the F\(_1\) layer, they acquire a center-of-mass momentum \(Q_{1}=2h_{1}/(\hbar v_{F})\). As a result, their wave function obtains an additional phase of \(Q_1 d_1\), leading to the following expression~\cite{Eschg,Eschrig}:
        \begin{equation}
            (\uparrow\downarrow)_{z}e^{iQ_{1}d_{1}}-(\downarrow\uparrow)_{z}e^{-iQ_{1}d_{1}}.
             \label{F1}
        \end{equation}
       When the magnetization in the F\(_{2}\) layer rotates by a certain angle (\(\theta\), \(\chi\)), the electrons forming the Cooper pairs experience a transformation upon entering the F\(_2\) layer~\cite{Eschg,Eschrig,HMXWYR2022},
      \begin{subequations}
             \begin{align}
               (\uparrow)_{z}\longrightarrow&(\uparrow)_{\theta,\chi}\cos\frac{\theta}{2}e^{i\chi/2}e^{i(k_{F}+Q_{2}/2)d_{2}}
                              -(\downarrow)_{\theta,\chi}\sin\frac{\theta}{2}e^{i\chi/2}e^{i(k_{F}-Q_{2}/2)d_{2}}, \label{Spa}\\
               (\downarrow)_{z}\longrightarrow&(\uparrow)_{\theta,\chi}\sin\frac{\theta}{2}e^{-i\chi/2}e^{i(k_{F}+Q_{2}/2)d_{2}}
                              +(\downarrow)_{\theta,\chi}\cos\frac{\theta}{2}e^{-i\chi/2}e^{i(k_{F}-Q_{2}/2)d_{2}}. \label{Spb}
             \end{align}
        \end{subequations}
        Correspondingly, the Cooper pairs have the following transformation:
            \begin{align}
               (\uparrow\downarrow)_{z}e^{iQ_{1}d_{1}}-(\downarrow\uparrow)_{z}e^{-iQ_{1}d_{1}}\longrightarrow &
               (\uparrow\downarrow-\downarrow\uparrow)_{\theta,\chi}(\cos{Q_{1}d_{1}}\cos{Q_{2}d_{2}}-\cos\theta\sin{Q_{1}d_{1}}\sin{Q_{2}d_{2}}) \nonumber \\
               +&i(\uparrow\downarrow+\downarrow\uparrow)_{\theta,\chi}(\cos\theta\sin{Q_{1}d_{1}}\cos{Q_{2}d_{2}}+\cos{Q_{1}d_{1}}\sin{Q_{2}d_{2}}) \nonumber \\
               +&i(\uparrow\uparrow-\downarrow\downarrow)_{\theta,\chi}\sin\theta\sin{Q_{1}d_{1}}.
               \label{Cpp}
            \end{align}
        When the magnetization of the F\(_2\) layer is parallel to the F\(_1\) layer (\(\theta=0\)), the transformation process can be simplified as
            \begin{align}
               (\uparrow\downarrow)_{z}e^{iQ_{1}d_{1}}-(\downarrow\uparrow)_{z}e^{-iQ_{1}d_{1}}\longrightarrow &
               (\uparrow\downarrow-\downarrow\uparrow)_{\theta,\chi}\cos{(Q_{1}d_{1}+Q_{2}d_{2})} \nonumber \\
               +&i(\uparrow\downarrow+\downarrow\uparrow)_{\theta,\chi}\sin{(Q_{1}d_{1}+Q_{2}d_{2})}.
               \label{pp}
            \end{align}
        The phases acquired by the Cooper pairs can be expressed as \(Q_{1}d_{1} = \left(\frac{2h_{1}}{\hbar v_F}\right) d_{1} = \left(\frac{h_1}{E_{F}}\right)(k_{F}d_{1})\) and \(Q_{2}d_{2} = \left(\frac{h_2}{E_{F}}\right)(k_{F}d_{2})\), respectively. In the absence of the potential barrier, the spin-singlet pairs \((\uparrow\downarrow - \downarrow\uparrow)\) make a significant contribution to the Josephson current, with their amplitude oscillating according to the function \(\cos\left[\left(\frac{h_1}{E_{F}}\right)(k_{F}d_{1}) + \left(\frac{h_2}{E_{F}}\right)(k_{F}d_{2})\right]\). As a result, the critical current oscillates as \( h_{1(2)}/E_{F} \) increases, exhibiting a period of \( 0.2 \), provided that \( k_{F}d_{1} = k_{F}d_{2} = 10\pi \) (see Fig.~\ref{Fig2}(a)). Similarly, when \( h_{1}/E_F = h_{2}/E_F = 0.1 \), the critical current oscillates with a period of \( 20\pi \) as a function of \( k_{F}d_{1(2)} \), as illustrated in Fig.~\ref{Fig2}(d).

        In contrast, resonant tunneling occurs when the potential barrier separates the F\(_1\) and F\(_2\) layers. As shown in Figs.~\ref{Fig2}(b) and \ref{Fig2}(e), the critical current displays regular periodic resonance peaks as the exchange field and ferromagnetic thickness increase. Specifically, when \( k_{F}d_{1} = k_{F}d_{2} = 10\pi \), these resonance peaks appear at positions \( h_{1(2)}/E_{F} = 0.1\left(n_{1(2)} + 1/2\right) \), where \( n_{1(2)} = 0, 1, 2, \cdots \) (see Fig.~\ref{Fig2}(b)). Moreover, as depicted in Fig.~\ref{Fig2}(e), the peak positions follow the relation \( k_{F}d_{1(2)} =10\pi\left(n_{1(2)} + 1/2\right) \) under the condition of \( h_{1}/E_{F} = h_{2}/E_{F} = 0.1 \). From the above two relations, we can derive a general resonance condition \( \left( h_{1(2)}/E_F \right) \left(k_{F}d_{1(2)} \right) \\ = \left( n_{1(2)} + 1/2 \right) \pi \). It can be simplified to \( Q_{1(2)}d_{1(2)} = \left( n_{1(2)} + 1/2 \right) \pi \). These results are due to the transport of the spin-triplet pairs \((\uparrow\downarrow + \downarrow\uparrow)\) within the ferromagnetic region. The spin-triplet pairs are odd in frequency and even in momentum, which makes them insensitive to nonmagnetic impurities~\cite{Bergeret}. Consequently, the potential barrier obstructs the transport of the spin-singlet pairs (\(\uparrow\downarrow-\downarrow\uparrow\)), thereby highlighting the contribution of the spin-triplet pairs (\(\uparrow\downarrow+\downarrow\uparrow\)) to the Josephson current. These spin-triplet pairs oscillate in a sinusoidal manner, represented by \(\sin(Q_{1}d_{1})\) in the F\(_{1}\) layer and \(\sin(Q_{2}d_{2})\) in the F\(_{2}\) layer. Thereupon, their amplitudes reach the maximum at \(Q_{1}d_{1} = (n_{1} + 1/2)\pi\) and \(Q_{2}d_{2} = (n_{2} + 1/2)\pi\). Under these conditions, the tunneling probability of the spin-triplet pairs is the largest, resulting in the highest amplitude of critical current. It is important to note that the heights of the current resonance peaks are nearly identical. However, the currents at different resonance peaks correspond to varying phases \(\phi_t =Q_{1}d_{1}+Q_{2}d_{2}=(n_{1} + n_{2} + 1)\pi\). These phases represent the total phase acquired by the spin-triplet pairs as they traverse through the F\(_{1}\) and F\(_{2}\) layers. The phase \(\phi_t\) ultimately determines the ground state of the Josephson junction. When \(\phi_t\) is an even number, the Josephson junction is in the 0-state. Conversely, when \(\phi_t\) is an odd number, it transitions to the \(\pi\)-state. Additionally, the resonant tunneling effect is significantly affected by temperature. As depicted in Figs.~\ref{Fig2}(c) and \ref{Fig2}(f), the critical current decreases and returns to its original stripe pattern at higher temperatures.

        \begin{figure}
          \centering
          \includegraphics[width=5.8in]{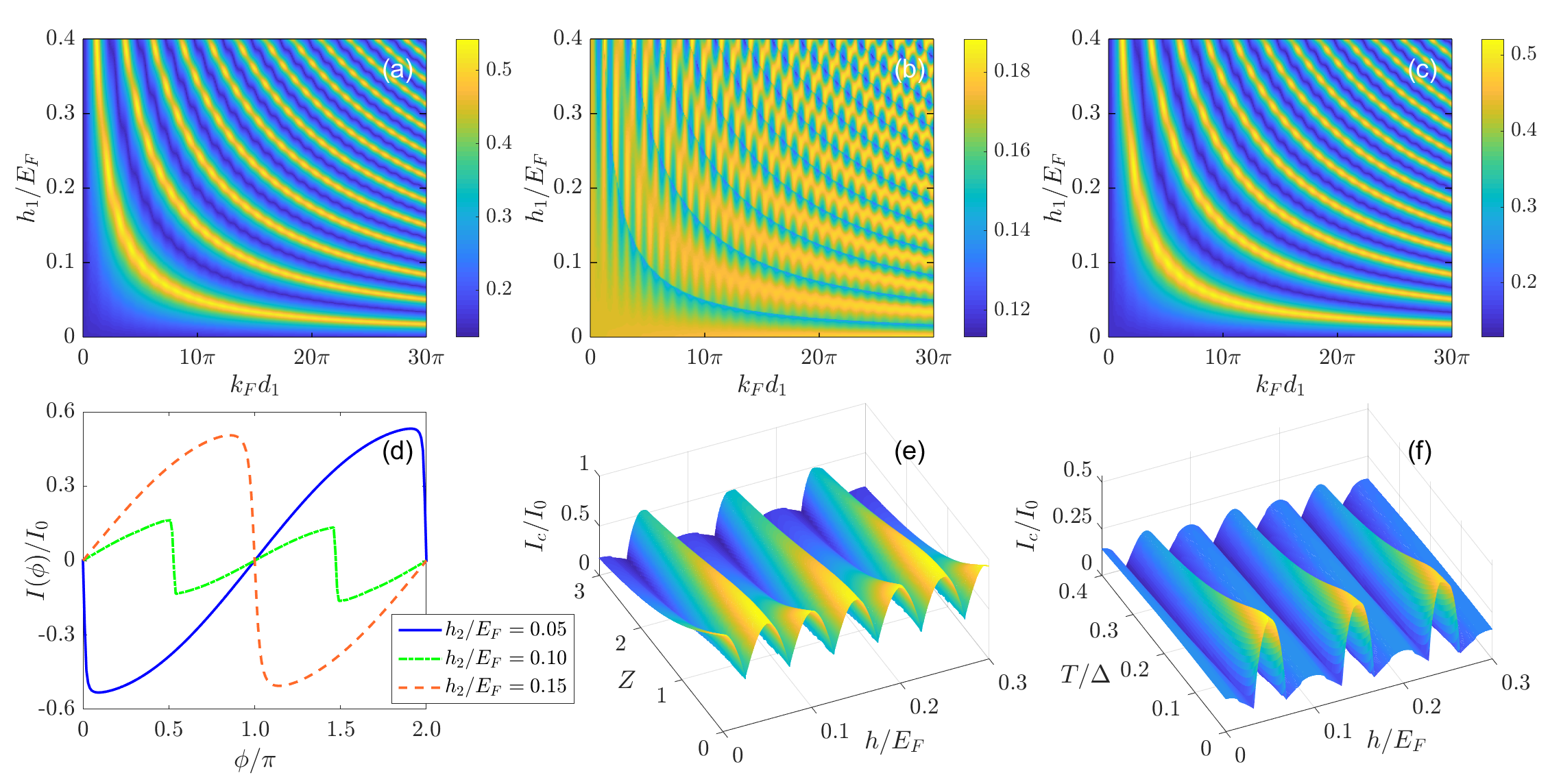} 
          \caption{The critical current \( I_{c} \) versus the exchange field \( h_1 \) and the thickness \( d_1 \) for a fixed thickness \( k_{F}d_{2} = 10\pi \) and different parameters [(a) \( h_{2}/E_{F} = 0.05 \), (b) \( h_{2}/E_{F} = 0.10 \), and (c) \( h_{2}/E_{F} = 0.15 \)]. (d) The current-phase relation \(I(\phi)\) for three different exchange fields $h_{2}$ with \( h_{1}/E_{F} = 0.05 \) and \( k_{F}d_{1} = k_{F}d_{2} = 10\pi \). (e) \( I_{c} \) versus the exchange field \( h \) and the barrier strength \( Z \). (f) \( I_{c} \) versus the exchange field \( h \) and temperature \( T \). For both panels (e) and (f), the parameters are set as \( h_{1} = h_{2} = h \) and \( k_{F}d_{1} = k_{F}d_{2} = 10\pi \). The barrier strength in all panels except (e) is \( Z = 3 \), and the temperature in all panels except (f) is \( T/\Delta = 0 \). All panels are for the parallel magnetization configurations (\( \theta = 0 \) and \( \chi = 0 \)).}
          \label{Fig3}
        \end{figure}

       We continue to analyze how the critical current varies with the exchange field \(h_1\) and the thickness \(d_1\) of the F\(_1\) layer for different F\(_2\) layers. Figure~\ref{Fig3}(a) illustrates that the critical current exhibits a rippled structure, with the two variables \((h_1/E_F, k_F d_1)\) or \((Q_1, d_1)\) showing an inversely proportional relationship. This ripple pattern results from the quantized resonant tunneling effect, where the discrete wave peaks correspond to distinct quantum states. In this scenario, the parameters of the F\(_2\) layer are fixed at \(h_2/E_F = 0.05\) and \(k_F d_2 = 10\pi\), corresponding to the resonance condition \(Q_2 d_2 = (n_2 + 1/2) \pi\) for \(n_2 = 0\). To achieve the resonant tunneling effect, the F\(_1\) layer must also meet the condition \(Q_1 d_1 = (n_1 + 1/2) \pi\). Consequently, the wave peaks from left to right correspond to the quantum numbers \(n_1 = 0, 1, 2, \cdots\). The total phase at these wave peaks is \(\phi_t = (n_1 + 1)\pi\). When the F\(_2\) layer deviates from the resonance condition--specifically, when \(h_2/E_F = 0.1\), leading to \(Q_2 d_2 = \pi\)--the critical current decreases, and the resonance peaks transform into troughs (see Fig.~\ref{Fig3}(b)). This phenomenon occurs because the amplitude of the spin-triplet pairs in the F\(_2\) layer depends on \(\sin(Q_2 d_2)\), which becomes zero at \(Q_2 d_2 = \pi\). As a result, the transmission of spin-triplet pairs is suppressed in the F\(_2\) layer, even though the F\(_1\) layer satisfies the resonant tunneling condition. When the exchange field in the F\(_2\) layer increases to \(h_2/E_F = 0.15\) (corresponding to \(Q_2 d_2 = 1.5\pi\)), the resonant tunneling pattern reemerges. As illustrated in Fig.~\ref{Fig3}(c), the characteristics of the critical current resemble those observed at \(h_2/E_F = 0.05\). However, the total phase at the wave peaks shifts to \(\phi_t = (n_1 + 2)\pi\), indicating that the phase \(\phi_t\) at the same wave peaks in Figs.~\ref{Fig3}(a) and \ref{Fig3}(c) differs by \(\pi\).

       To further clarify this issue, we illustrate the current-phase relation under three different exchange fields \(h_{2}\) in Fig.~\ref{Fig3}(d). With the parameters \(h_{1}/E_{F}=0.05\) and \(k_{F}d_{1}=10\pi\), the spin-triplet pairs acquire a phase of \(Q_{1}d_{1} = 0.5\pi\) as they pass through the F\(_1\) layer. Additionally, when they cross the F\(_2\) layer, they gain an extra phase of \(Q_{2}d_{2} = 0.5\pi\), assuming that the F\(_2\) layer has the same parameters (\(h_{2}/E_{F}=0.05\) and \(k_{F}d_{2}=10\pi\)). In this situation, the Josephson junction remains in the \(\pi\)-state because the total phase gained by these spin-triplet pairs is \(\phi_t = \pi\). However, when the exchange field increases to \(h_{2}/E_{F}=0.1\), the spin-triplet pairs obtain a phase of \(Q_{2}d_{2} = \pi\) in the F\(_2\) layer. This results in their amplitude decreasing to zero, since \(\sin(Q_{2}d_{2}) = 0\). Hence, the first harmonic current (\(I_{1}\)) vanishes, leaving only the second harmonic current (\(I_{2}\)). Furthermore, if the exchange field \(h_{2}/E_{F}\) rises to \(0.15\), the phase acquired by the spin-triplet pairs becomes \(Q_{2}d_{2} = 1.5\pi\). In this case, the Josephson junction transitions to the 0-state because \(\phi_t = 2\pi\). These analytical results are consistent with the current-phase relation depicted in Fig.~\ref{Fig3}(d).

       When two ferromagnetic layers have identical exchange fields and thicknesses, the behavior of the critical current becomes quite intriguing. As shown in Fig. \ref{Fig3}(e), in the absence of the potential barrier, the critical current oscillates periodically with increasing exchange field. This oscillation is the characteristic of the 0-\(\pi\) transitions driven by the spin-singlet pairs. However, if the potential barrier is sufficiently large (for instance, \(Z=3\)), the oscillation peaks associated with the 0-state nearly disappear, while those corresponding to the \(\pi\)-state remain pronounced. This selection behavior arises from the resonant tunneling of the spin-triplet pairs, which occurs when the condition \(Q_{1}d_{1}=Q_{2}d_{2}=(n_{1}+1/2)\pi\) is met. Under this condition, the total phase accumulation for the spin-triplet pairs becomes \(\phi_{t}=(2n_{1}+1)\pi\), effectively locking the junction into the \(\pi\)-state. Consequently, the potential barrier acts as a filter, allowing current transport in the \(\pi\)-state while suppressing that in the 0-state. Furthermore, the resonant tunneling current diminishes as the temperature rises. As shown in Fig.~\ref{Fig3}(f), at higher temperatures, the original 0-\(\pi\) oscillation pattern reemerges as thermal decoherence overcomes the resonant tunneling mechanism.

       \subsection{The Josephson current in configurations with perpendicular magnetization}

        \begin{figure}[ptb]
           \centering
           \includegraphics[width=5.8in]{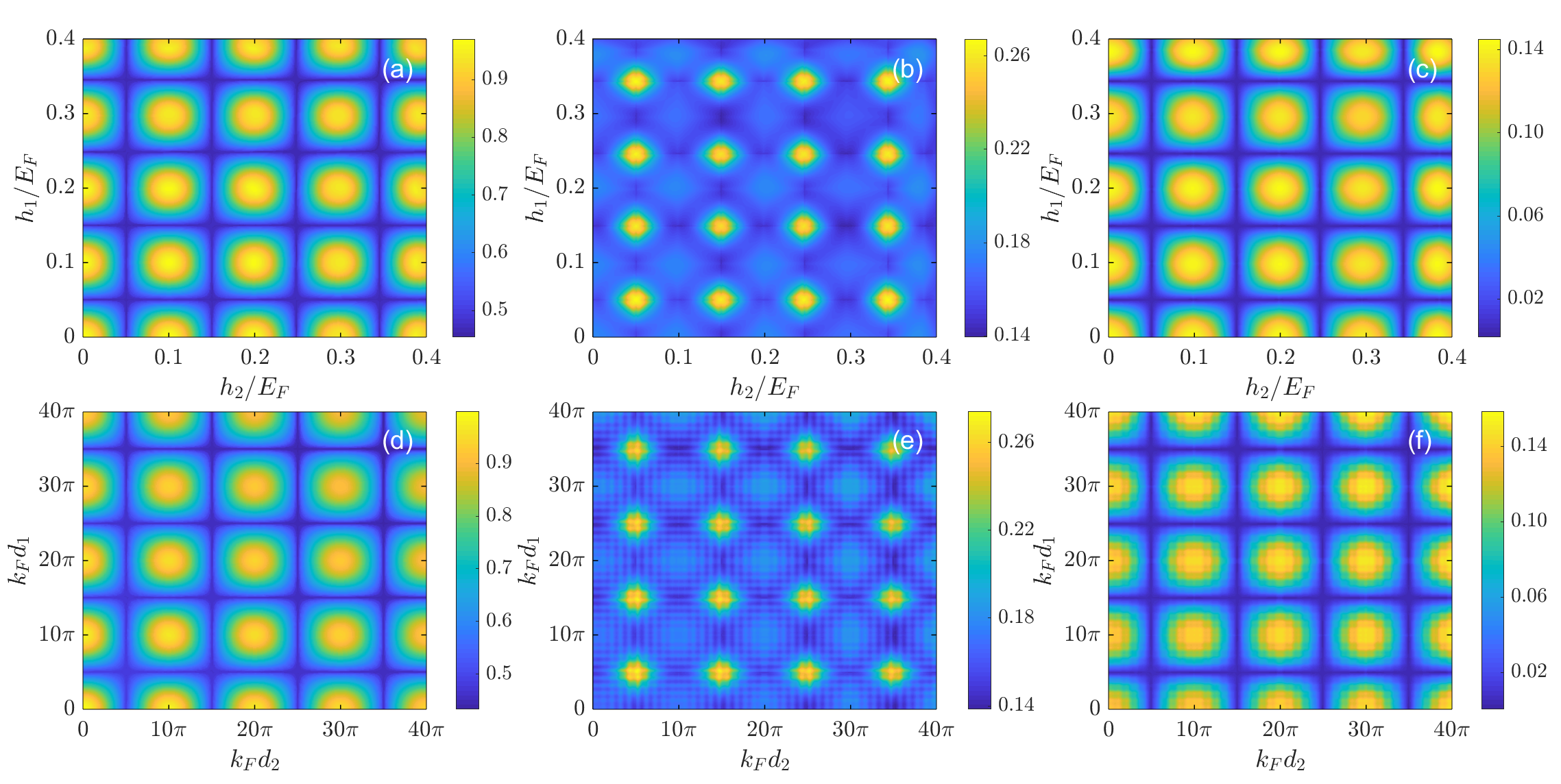}
           \caption{The critical current \( I_{c} \) versus the exchange fields (\( h_1 \), \( h_2 \)) for the thicknesses \( k_{F}d_{1} = k_{F}d_{2} = 10\pi \) [(a), (b), and (c)], and \( I_{c} \) versus (\( d_{1} \), \( d_{2} \)) for \( h_{1}/E_F = h_{2}/E_F = 0.1 \) [(d), (e), and (f)]. The left column of graphs [(a) and (d)] corresponds to the barrier strength \( Z = 0 \), and the middle [(b) and (e)] and right [(c) and (f)] columns correspond to \( Z = 3 \). Additionally, the temperature is taken as \( T/\Delta = 0 \) for the left [(a) and (d)] and middle [(b) and (e)] columns and \( T/\Delta = 0.4 \) for the right [(c) and (f)] column. All panels are for the perpendicular magnetization configurations (\( \theta = \pi/2 \) and \( \chi = \pi/2 \)).}
           \label{Fig4}
        \end{figure}

      In the configurations with perpendicular magnetization, the critical current exhibits notable variations. As illustrated in Figs.~\ref{Fig4}(a) and \ref{Fig4}(d), when the potential barrier is absent, the current peaks are arranged periodically in a ``square lattice'' as the exchange fields (\(h_{1}\), \(h_{2}\)) and the ferromagnetic thicknesses (\(d_{1}\), \(d_{2}\)) change. This pattern arises from the phase accumulation of the spin-singlet pairs in the ferromagnetic layers. For the perpendicular case (\(\theta = \pi/2\)), the transformation process of the Cooper pairs, described by the formula (\ref{Cpp}), can be simplified to
            \begin{align}
               (\uparrow\downarrow)_{z}e^{iQ_{1}d_{1}}-(\downarrow\uparrow)_{z}e^{-iQ_{1}d_{1}}\longrightarrow &
               (\uparrow\downarrow-\downarrow\uparrow)_{\pi/2,\chi}\cos{Q_{1}d_{1}}\cos{Q_{2}d_{2}}  \nonumber \\
               +&i(\uparrow\downarrow+\downarrow\uparrow)_{\pi/2,\chi}\cos{Q_{1}d_{1}}\sin{Q_{2}d_{2}} \nonumber \\
               +&i(\uparrow\uparrow-\downarrow\downarrow)_{\pi/2,\chi}\sin{Q_{1}d_{1}}. \label{CppCZ}
            \end{align}

          \begin{figure}[ptb]
             \centering
             \includegraphics[width=5.8in]{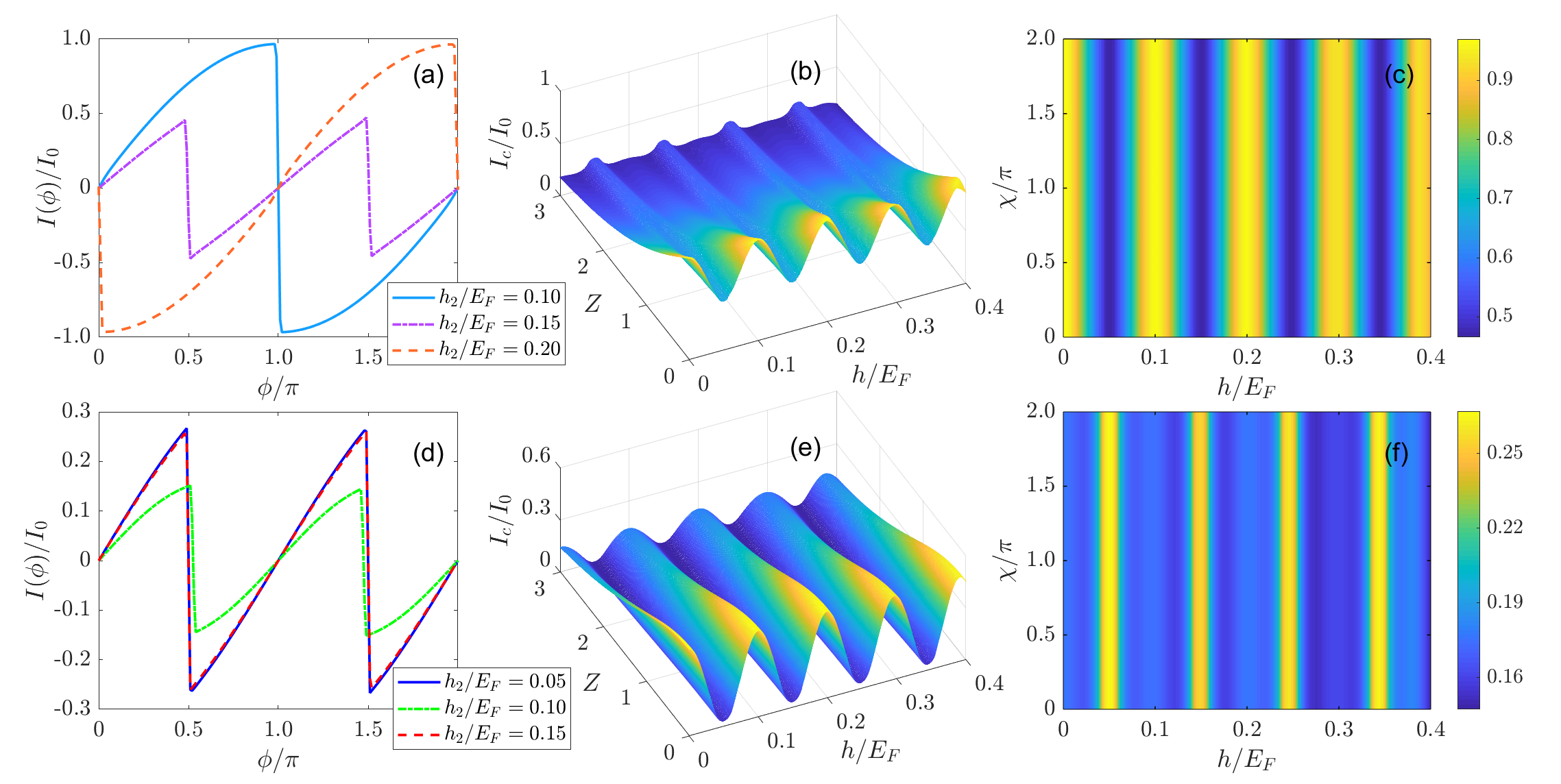}
             \caption{The current-phase relation \(I(\phi)\) for three values of \( h_2 \) in the case of (\( h_1/E_F = 0.1 \), \( Z = 0 \)) (a) and (\( h_1/E_F = 0.05 \), \( Z = 3 \)) (d), when the temperature is \( T/\Delta = 0 \). The critical current \( I_c \) versus (\( h, Z \)) for \( T/\Delta = 0 \) (b) and \( T/\Delta = 0.4 \) (e). In panels [(a), (b), (d), and (e)], the results are for the perpendicular magnetization configurations (\( \theta = \pi/2 \) and \( \chi = \pi/2 \)). \( I_c \) versus \( (h, \chi) \) for \( Z = 0 \) (c) and \( Z = 3 \) (f) when \( T/\Delta = 0 \) and \(\theta=\pi/2\). In panels [(b), (c), (e), and (f)], we define \( h_1 = h_2 = h \). In all panels, the ferromagnetic thicknesses are taken as \( k_F d_1 = k_F d_2 = 10\pi \).}
             \label{Fig5}
          \end{figure}

      The amplitude of the spin-singlet pairs \((\uparrow\downarrow - \downarrow\uparrow)\) varies as \(\cos(Q_{1}d_{1})\cos(Q_{2}d_{2})\), resulting in fluctuations of the critical current. Therefore, the critical current reaches its maximum when \(Q_{1(2)}d_{1(2)} = n_{1(2)}\pi\) and its minimum when \(Q_{1}d_{1} = (n_{1} + 1/2)\pi\) or \(Q_{2}d_{2} = (n_{2} + 1/2)\pi\). To clarify these findings, we present the current-phase relation in Fig.~\ref{Fig5}(a) for three specific exchange fields \(h_2\) while keeping the parameters fixed at \(h_{1}/E_{F} = 0.1\) and \(k_{F}d_{1} = k_{F}d_{2} = 10\pi\). Under these conditions, the spin-singlet pairs acquire a phase of \(Q_1d_1 = \pi\) as they cross the F\(_1\) layer. When the exchange field in the F\(_2\) layer is \(h_{2}/E_{F} = 0.1\), the spin-singlet pairs also gain a phase of \(Q_2d_2 = \pi\) after passing through the F\(_2\) layer, which places the Josephson junction in the 0-state. As the exchange field \(h_{2}/E_{F}\) increases to 0.15, the phase changes to \(Q_2d_2 = 1.5\pi\), resulting in the disappearance of the first harmonic current (\(I_{1}\)) and the emergence of the second harmonic current (\(I_{2}\)). When the exchange field \(h_{2}/E_{F}\) increases further to 0.2, at which point \(Q_2d_2 = 2\pi\), the Josephson junction transitions into the \(\pi\)-state.

      In contrast, the introduction of the potential barrier at the F\(_1\)/F\(_2\) interface results in the formation of current resonance peaks at \( Q_{1}d_{1} = (n_{1} + 1/2)\pi \) and \( Q_{2}d_{2} = (n_{2} + 1/2)\pi \), as illustrated in Figs. \ref{Fig4}(b) and \ref{Fig4}(e). This behavior arises because the potential barrier filters out the spin-singlet pairs \( (\uparrow\downarrow - \downarrow\uparrow) \), leading to the disappearance of the first harmonic current (\( I_{1} \)). As a result, the remaining spin-triplet pairs manifest as \( (\uparrow\downarrow + \downarrow\uparrow)_{z} \) in the F\(_1\) layer and transform into \( (\uparrow\uparrow - \downarrow\downarrow)_{\pi/2, \chi} \) in the F\(_2\) layer. This phenomenon resembles the resonant tunneling of two entangled spin-triplet pairs [\( (\uparrow\downarrow)_{z}\) and \((\downarrow\uparrow)_{z}\)] or [\( (\uparrow\uparrow)_{\pi/2, \chi}\) and \((\downarrow\downarrow)_{\pi/2, \chi}\)] in the ferromagnetic regions, enabling the full emergence of the second harmonic current (\( I_{2} \))~\cite{LTrifunov,Richard,HMeng}. To illustrate these findings, Fig. \ref{Fig5}(d) presents the current-phase relations for three specific exchange fields \( h_{2}/E_{F} \), with fixed parameters of \( h_{1}/E_{F} = 0.05 \) and \( k_{F}d_{1} = k_{F}d_{2} = 10\pi \). Notably, at \( h_{2}/E_{F} = 0.05 \) and \( 0.15 \), the current amplitudes are significant, corresponding to the resonance peak of the critical current. In contrast, at \( h_{2}/E_{F} = 1.0 \), the current amplitude is diminished, aligning with the trough of the critical current. In all cases, the period of \( I(\phi) \) concerning \(\phi\) is \( \pi \),  which is a hallmark of the second harmonic current (\( I_{2} \)). Moreover, even in the presence of the potential barrier, elevated temperatures can suppress the resonant tunneling effect. As illustrated in Figs. 4(c) and 4(f), when the barrier strength is \( Z = 3 \) and the temperature at \( T/\Delta = 0.4 \), the variations in the critical current closely resemble those observed in the junctions without potential barriers (as depicted in Figs. \ref{Fig4}(a) and \ref{Fig4}(d)). The main difference is a decrease in the current amplitude.

         Next, we explore the dependence of the critical current on the barrier strength \( Z \) at various temperatures. As depicted in Fig. \ref{Fig5}(b), at zero temperature (\( T/\Delta = 0 \)), the amplitude of the critical current decreases as the barrier strength \( Z \) increases, with transitions where the peaks switch to valleys and vice versa. In contrast, at a higher temperature (\( T/\Delta = 0.4 \)), the critical current monotonically decreases with increasing barrier strength \( Z \), and there are no transitions between peaks and valleys, as illustrated in Fig.~\ref{Fig5}(e). Figures~\ref{Fig5}(c) and \ref{Fig5}(f) show that the critical current oscillates periodically with the exchange field \( h \) (where \( h_{1} = h_{2} = h \)) and is unaffected by the azimuthal angle \( \chi \). In the absence of the potential barrier (\( Z = 0 \)), the critical current reaches its peak values at \( h/E_{F} = 0 \), 0.1, 0.2, 0.3, and 0.4. These peaks correspond to conditions where \( Q_{1}d_{1} = Q_{2}d_{2} = 0 \), \( \pi \), \( 2\pi \), \( 3\pi \), and \( 4\pi \). However, upon introducing a finite potential barrier (\(Z = 3\)), the resonance peaks shift to \(h/E_{F} = 0.05\), 0.15, 0.25, and 0.35, corresponding to \(Q_{1}d_{1} = Q_{2}d_{2} = 0.5\pi\), \(1.5\pi\), \(2.5\pi\), and \(3.5\pi\). These observations suggest that the potential barrier can shift the positions of the resonance peaks of the critical current.

         \subsection{The Josephson current in configurations with antiparallel magnetization}

        \begin{figure}[ptb]
          \centering
          \includegraphics[width=5.8in]{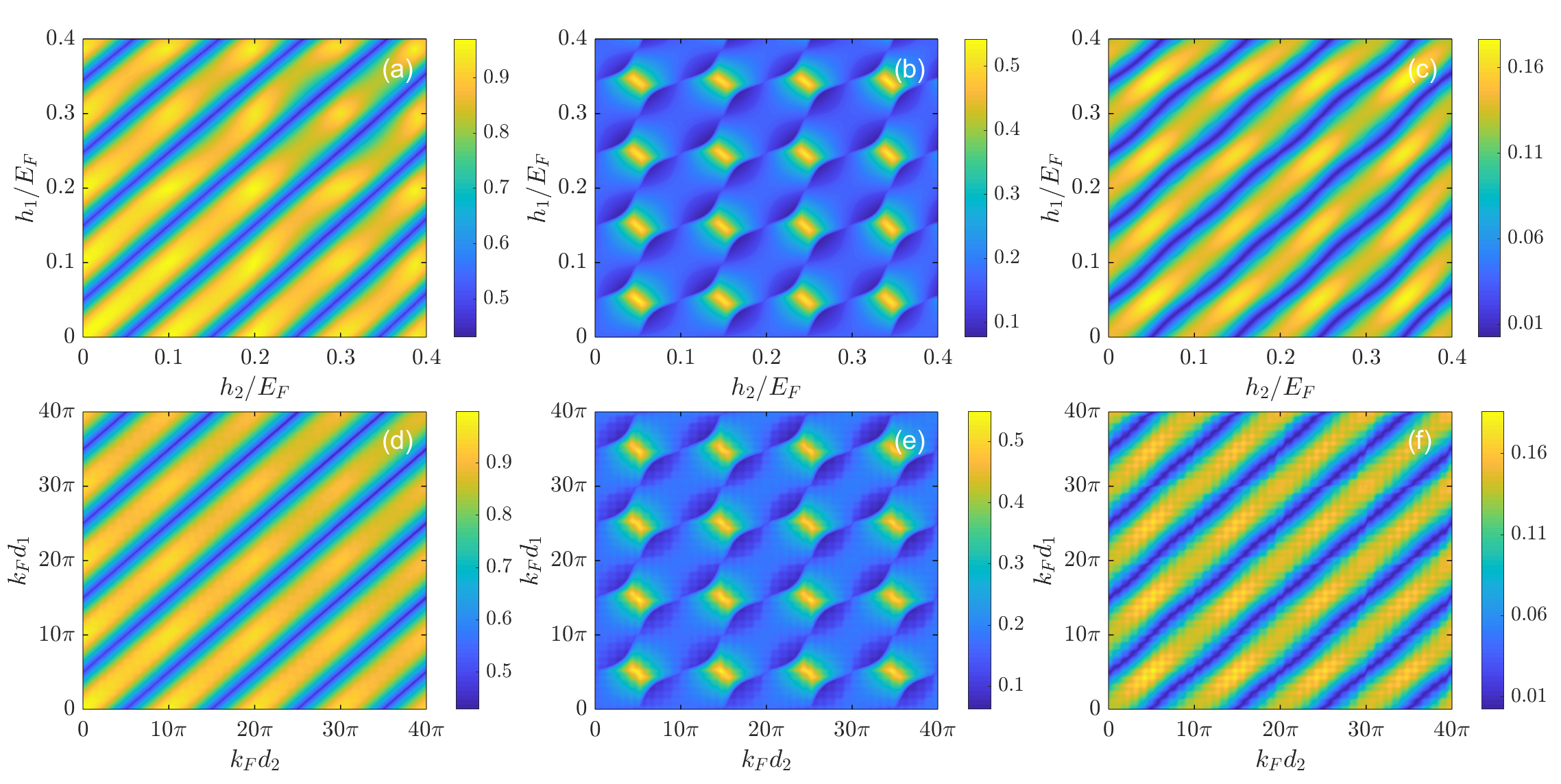}
          \caption{The critical current \( I_{c} \) versus the exchange fields (\( h_1 \), \( h_2 \)) for the thicknesses \( k_{F}d_{1} = k_{F}d_{2} = 10\pi \) [(a), (b), and (c)], and \( I_{c} \) versus (\( d_{1} \), \( d_{2} \)) for \( h_{1}/E_F = h_{2}/E_F = 0.1 \) [(d), (e), and (f)]. The left column of graphs [(a) and (d)] corresponds to the barrier strength \( Z = 0 \), and the middle [(b) and (e)] and right [(c) and (f)] columns correspond to \( Z = 3 \). Additionally, the temperature is taken as \( T/\Delta = 0 \) for the left [(a) and (d)] and middle [(b) and (e)] columns and \( T/\Delta = 0.4 \) for the right [(c) and (f)] column. All panels are for the antiparallel magnetization configurations (\( \theta = \pi \) and \( \chi = 0 \)).}
          \label{Fig6}
        \end{figure}

        As illustrated in Figs.~\ref{Fig6}(a) and \ref{Fig6}(d), in antiparallel magnetization configurations, the critical current shows continuous oscillating stripes as the exchange fields (\(h_{1}\), \(h_{2}\)) and the ferromagnetic thicknesses (\(d_{1}\), \(d_{2}\)) vary, provided no potential barrier is present. These stripes have a positive slope, which differs from the behavior observed in parallel configurations. The critical current oscillates concerning the exchange field \(h_{1(2)}/E_F\) with a period of 0.2, while its oscillation period related to the ferromagnetic thickness \(k_Fd_{1(2)}\) is \(20\pi\). To clarify this mechanism, we analyze the evolution of the Cooper pair wave function using the formula (\ref{Cpp}). In the case of antiparallel alignment (\(\theta = \pi\)), as the Cooper pairs move through the F\(_1\) and F\(_2\) layers, they undergo a transformation process:
          \begin{align}
               (\uparrow\downarrow)_{z}e^{iQ_{1}d_{1}}-(\downarrow\uparrow)_{z}e^{-iQ_{1}d_{1}}\longrightarrow &
               (\uparrow\downarrow-\downarrow\uparrow)_{\pi,\chi}\cos({Q_{1}d_{1}-Q_{2}d_{2}})  \nonumber \\
               -&i(\uparrow\downarrow+\downarrow\uparrow)_{\pi,\chi}\sin({Q_{1}d_{1}-Q_{2}d_{2}}). \label{CppCZ}
          \end{align}
         Without the potential barrier, the spin-singlet pairs \((\uparrow\downarrow - \downarrow\uparrow)\) contribute significantly to the Josephson current. The amplitude of these spin-singlet pairs is proportional to the cosine function \(\cos(Q_{1}d_{1} - Q_{2}d_{2})\). Consequently, the critical current experiences periodic oscillations based on the variables \(Q_{1}d_{1}\) and \(Q_{2}d_{2}\).

        In contrast, as shown in Figs.~\ref{Fig6}(b) and \ref{Fig6}(e), the presence of the potential barrier results in periodic resonance peaks in the critical current. Notably, these peaks appear at the same positions as those found in the configurations with parallel magnetizations. This phenomenon occurs because the spin-singlet pairs \((\uparrow\downarrow - \downarrow\uparrow)\) are suppressed, while the spin-triplet pairs \((\uparrow\downarrow + \downarrow\uparrow)\) play a crucial role in generating the Josephson current. The spin-triplet pairs reach their maximum amplitude at \(Q_{1}d_{1} = (n_{1} + 1/2)\pi\) and \(Q_{2}d_{2} = (n_{2} + 1/2)\pi\), causing the critical current to achieve its resonance peaks at these specific points. At these resonance peaks, the total phase acquired by the spin-triplet pairs is \(\phi_{t} = Q_{1}d_{1} - Q_{2}d_{2}=(n_{1} - n_{2})\pi\), which differs from that in the configurations with parallel magnetization. It is also important to note that higher temperatures suppress the resonant tunneling effect. For instance, at \(T/\Delta = 0.4\), the critical current diminishes and returns to the original periodic stripe pattern, as depicted in Figs. \ref{Fig6}(c) and \ref{Fig6}(f).

        \begin{figure}[ptb]
           \centering
           \includegraphics[width=5.8in]{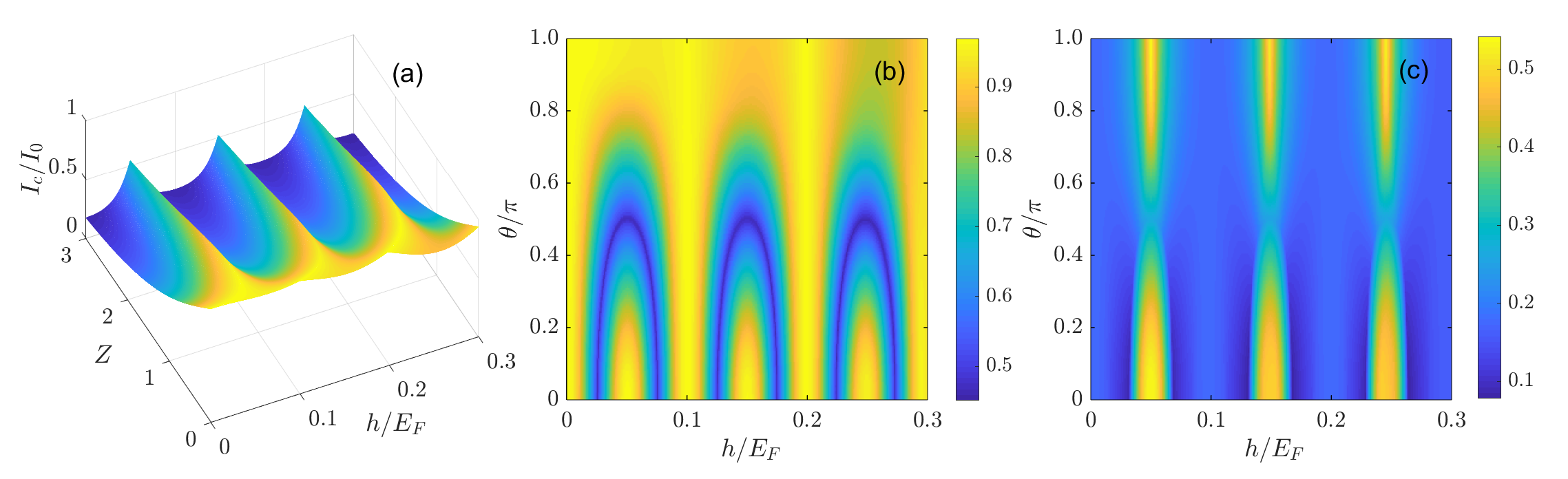}
           \caption{The critical current \( I_{c} \) versus the exchange fields \( h \) and the barrier strength \( Z \) for the antiparallel magnetization configurations (\( \theta=\pi \) and \( \chi=0 \)) (a), and \( I_{c} \) versus (\( h \), \( \theta \)) for \( Z=0 \) (b) and \( Z=3 \) (c), where the ferromagnetic layers have the same exchange field \( h_1=h_2=h \). In panels [(b) and (c)], the azimuthal angle is set to \( \chi = \pi/2 \). For all panels, the ferromagnetic thicknesses and temperature are \( k_{F}d_{1} = k_{F}d_{2} = 10\pi \) and \( T/\Delta = 0 \), respectively.}
           \label{Fig7}
        \end{figure}

       A particularly striking scenario occurs when two ferromagnetic layers have identical properties, specifically equal exchange fields (\( h_1 = h_2 \)) and equal thicknesses (\( d_1 = d_2 \)), as illustrated in Fig.~\ref{Fig7}(a). In the absence of the potential barrier (\( Z = 0 \)), the antiparallel magnetizations completely cancel each other out, causing the entire ferromagnet to behave like a normal-metal. As a result, the critical current remains almost unchanged with the increase of exchange field or ferromagnetic thickness. However, as the potential barrier increases, the resonant tunneling effect becomes more pronounced, leading to the emergence of resonance peaks at \( Q_{1}d_{1} = (n_{1} + 1/2)\pi \) and \( Q_{2}d_{2} = (n_{2} + 1/2)\pi \), where \( n_{1} = n_{2} \). In this situation, the spin-triplet pairs \((\uparrow\downarrow + \downarrow\uparrow)\) acquire a total phase of \( \phi_{t} = Q_{1}d_{1} - Q_{2}d_{2} = 0 \). Consequently, the Josephson junction remains in the 0-state at these current resonance peaks.

       In Fig.~\ref{Fig7}(b), the critical current exhibits a periodic ``candle flame-like'' pattern when there is no potential barrier. In the range of \(0 < \theta < 0.5\pi\), the critical current oscillates as the exchange field increases, which is a manifestation of the 0-\(\pi\) transition. However, in the region of \(0.5\pi < \theta < \pi\), these oscillations gradually diminish with increasing \(\theta\). Notably, when \(\theta = \pi\) (indicating antiparallel alignment), the current amplitude reaches its maximum, and the oscillatory behavior nearly disappears. It is essential to highlight that as the magnetization of the two ferromagnets rotates from a parallel to an antiparallel alignment, the Josephson current exhibits two distinct behaviors under different exchange fields: (i) For \(h/E_{F} =\) 0, 0.1, 0.2, and \(0.3\), the critical current always keeps a considerable value with increasing \(\theta\), in which case the Josephson junction is in the 0-state. (ii) For \(h/E_F =\) 0.05, 0.15, and \(0.25\), the critical current initially decreases before rising again as \(\theta\) increases. This behavior reflects a transition from the \(\pi\)-state to the 0-state. In contrast, the potential barrier shows specific current selectivity. As illustrated in Fig.~\ref{Fig7}(c), the barrier blocks current in case (i) while allowing it to pass in case (ii). Thus, by adjusting the relative directions of magnetization between parallel and antiparallel arrangements, the junction can effectively switch between the 0- and \(\pi\)-states.

       \section{Conclusion}
       \label{Sec4}

      We have studied the Josephson current in the SF\(_{1}\)F\(_{2}\)S junctions with a tunable potential barrier at the F\(_{1}\)/F\(_{2}\) interface, utilizing exact numerical solutions of the Bogoliubov-de Gennes equations. Our findings reveal various phenomena across three distinct magnetization configurations, offering insights into fundamental physics and potential applications. The key results are summarized as follows:

     (i) In parallel magnetization configurations, the critical current oscillates continuously with the exchange fields (\(h_1\), \(h_2\)) and the ferromagnetic thicknesses (\(d_1\), \(d_2\)) when the potential barrier is absent. This behavior results from the oscillations of the spin-singlet pairs (\(\uparrow\downarrow - \downarrow\uparrow\)) within the ferromagnetic layers. In this scenario, the amplitude of these oscillations is proportional to the function \(\cos(Q_1d_1 + Q_2d_2\)), where \(Q_1 d_1\) and \(Q_2 d_2\) take on continuous values. However, when the potential barrier exists, the critical current displays discrete periodic resonance peaks under quantization conditions \(Q_1 d_1 = (n_1 + 1/2)\pi\) and \(Q_2 d_2 = (n_2 + 1/2)\pi\) at low temperatures. These resonance peaks arise from the resonant tunneling of the spin-triplet pairs (\(\uparrow\downarrow + \downarrow\uparrow\)). As these pairs traverse the F\(_1\) and F\(_2\) layers, they accumulate a total phase \(\phi_t = Q_1 d_1 + Q_2 d_2 = (n_1 + n_2 + 1)\pi\), which determines the ground state of the Josephson junction. Additionally, as the temperature increases, the resonant tunneling effect diminishes, causing the resonance peaks to disappear and the original oscillatory behavior to resurface.

     (ii) In perpendicular magnetization configurations, the critical current depends on the function \(\cos(Q_1 d_1)\cos(Q_2 d_2)\) when there is no potential barrier present. However, the potential barrier selectively filters out the first harmonic current (\(I_1\)) and allows only the second harmonic current (\(I_2\)) to pass while still maintaining the quantized tunneling conditions mentioned above.

     (iii) In antiparallel magnetization configurations, the critical current oscillates according to \(\cos(Q_1d_1 - Q_2d_2)\) when no potential barrier is present. With the potential barrier, the resonance peaks still appear under the same tunneling conditions, but the total phase acquired by the spin-triplet pairs changes to \(\phi_t = Q_1d_1 - Q_2d_2 = (n_1 - n_2)\pi\).

     More importantly, if both ferromagnetic layers have identical properties, such as the same exchange fields and thicknesses, the quantum numbers will also be equal (\(n_1 = n_2\)). The resonance peaks induced by the potential barrier correspond to different ground states: the \(\pi\)-state for the parallel configuration and the 0-state for the antiparallel configuration. This relationship creates a direct pathway for achieving 0-\(\pi\) transitions by modulating the relative orientation of magnetization, which is essential for superconducting phase engineering. Our findings suggest that the SF\(_{1}\)F\(_{2}\)S junctions with interfacial barriers offer a versatile platform for controlling Josephson currents through resonant tunneling and magnetization arrangement. These systems have considerable potential for applications in superconducting spintronics and cryogenic memory.

    \section*{Acknowledgements}
    \paragraph{Funding information}
     This work was supported by the National Natural Science Foundation of China (Grant No.12174238), the Natural Science Basic Research Program of Shaanxi Province (Program No.2020JM-597), the Innovation Team of Shaanxi Universities (Grant No. 2022-94), and the School-level Youth Innovation Team of Shaanxi University of Technology.






\bibliography{SciPost_Example_BiBTeX_File.bib}


\end{document}